\documentclass[twocolumn]{aastex631}

\usepackage{newtxtext,newtxmath}
\usepackage[T1]{fontenc}

\DeclareRobustCommand{\VAN}[3]{#2}
\let\VANthebibliography\thebibliography
\def\thebibliography{\DeclareRobustCommand{\VAN}[3]{##3}\VANthebibliography}

\usepackage{float}
\usepackage{graphicx}	
\usepackage{amsmath}	
\usepackage{wasysym}           
\usepackage{graphicx}
\usepackage{epstopdf}
\usepackage{mathrsfs}
\usepackage{anyfontsize}
\usepackage{natbib}
\usepackage{color}
\usepackage{lipsum}
\usepackage{diagbox}
\usepackage{xcolor}

\begin{document}

\title{Stars Crushed by Black Holes. III. Mild Compression of Radiative Stars by Supermassive Black Holes}
\shorttitle{Stars Crushed by Black Holes III}



\author[0000-0001-6118-0023]{Suman Kumar Kundu}
\affiliation{Department of Physics, Syracuse University,
Syracuse, NY 13244, USA; skundu@syr.edu, ecoughli@syr.edu}

\author[0000-0003-3765-6401]{Eric R.~Coughlin}
\affiliation{Department of Physics, Syracuse University,
Syracuse, NY 13244, USA; skundu@syr.edu, ecoughli@syr.edu}

\author[0000-0002-2137-4146]{C.~J.~Nixon}
\affiliation{Department of Physics and Astronomy, University of Leicester,
Leicester, LE1 7RH, UK}
\shortauthors{Kundu, Coughlin, \& Nixon}
\begin{abstract}
A tidal disruption event (TDE) occurs when the gravitational field of a supermassive black hole (SMBH) destroys a star. For TDEs in which the star enters deep within the tidal radius, such that the ratio of the tidal radius to the pericenter distance  $\beta$ satisfies $\beta \gg 1$, the star is tidally compressed and heated. It was predicted that the maximum density and temperature attained during deep TDEs scale as $\propto \beta^3$ and $\propto \beta^2$, respectively, and nuclear detonation triggered by $\beta \gtrsim 5$, but these predictions have been debated over the last four decades. We perform Newtonian smoothed-particle hydrodynamics (SPH) simulations of deep TDEs between a Sun-like star and a $10^6 M_\odot$ SMBH for $2 \le \beta \le 10$. We find that neither the maximum density nor temperature follow the $\propto \beta^3$ and $\propto \beta^2$ scalings or, for that matter, any power-law dependence, and that the maximum-achieved density and temperature are reduced by $\sim$ an order of magnitude compared to past predictions. We also perform simulations in the Schwarzschild metric, and find that relativistic effects modestly increase the maximum density (by a factor of $\lesssim 1.5$) and induce a time lag relative to the Newtonian simulations, which is induced by time dilation. We also confirm that the time the star spends at high density and temperature is a very small fraction of its dynamical time. We therefore predict that the amount of nuclear burning achieved by radiative stars during deep TDEs is minimal.
\end{abstract}

\keywords{Astrophysical black holes (98) --- Black hole physics (159) --- Hydrodynamical simulations (767) --- Hydrodynamics (1963)--- Supermassive black holes (1663) --- Tidal disruption (1696)}

\section{Introduction}
\label{sec:intro}
 Tidal disruption events occur when a star gets so close to a supermassive black hole (SMBH) that the tides imparted by the latter tear the star apart \citep{Hill1975,Frank1976,Young1977,Hill78,Frank1978,Kato1978}. The observation of these events has received an impetus in the last decade, and current (e.g., Chandra, Swift, SRG/eROSITA) and upcoming high-cadence wide-field all-sky surveys (e.g. SKA, LSST, Einstein probe) promise an exciting time ahead (see \citealt{Gezari21} for a detailed review of the observational status). 

 The outcome of a TDE depends largely on how close the star comes to the SMBH, which is implicitly defined via $\beta \equiv r_{\rm t}/r_{\rm p}$, where $r_{\rm t} \equiv R_{\star}\left(M_{\bullet}/M_{\star}\right)^{1/3}$, the tidal radius, is roughly the distance at which the SMBH tidal force equals the self-gravity of the star of radius $R_{\star}$ and mass $M_{\star}$, and $r_{\rm p}$ is the point of closest approach between the two bodies \citep{Hill1975}. Events with $\beta \lesssim 1$ are partial TDEs, where a fraction of the star survives the encounter intact \citep{Guillochon13,Mainetti17,Coughlin19,Miles20}. In contrast, in events with $\beta \gg 1$ ---  ``deep TDEs'' --- the gravity of the SMBH overwhelms the self-gravity of the star and the star is compressed by the vertical component of the tidal field of the SMBH \citep{Laguna1993,Brassart08,Evans15,Tejeda17,Darbha19,Nixon22}. The degree of tidal compression suffered by a star in deep TDEs has been studied by, e.g., \citet{Wheeler71, Hill78,Lidskii79,Carter1982,Carter83,Carter85,Luminet86, Brassart08, Stone2013, Gafton2019}. 
 
 \citet[hereafter CL82]{Carter1982} and \citet[hereafter CL83]{Carter83} found that as the $\beta$ of the encounter increases, the star experiences an increasing degree of adiabatic compression and its density increases to a maximum value $\rho_{\rm max}$ at roughly the time it reaches the pericenter. For radiative stars modeled with the Eddington standard model (e.g., \citealt{hansen2012}), in events with $\beta \gtrsim 5$ they claimed $\rho_{\rm max}/\rho_{\rm c} = 0.22 \beta^3$, where $\rho_{\rm c}$ is the original, central stellar density. From Figure $13$ of \citet [LC86, hereafter]{Luminet86}, for a 3 $M_\odot$ standard-model star, the central density (temperature) increases by a factor of $\sim$ 50 (10) for $\beta=5$ and $\sim$ 500 (50) when $\beta=10$. As a consequence of this sharp increase in central density and temperature, these authors predicted that in $\beta \gtrsim 5$ encounters the energy released from the triple-$\alpha$ process ignites helium-burning reactions, which was supported by \cite{Pichon85,LP89b,LP89} and even a second burst of nuclear energy release was postulated \citep{LM85}. Almost immediate criticism of the work of Carter and Luminet came from \cite{Bicknell1983}, who used numerical techniques to refute the possibility of helium detonation as they found significantly milder compression. 
 Despite many efforts, to date the degree of tidal compression in deep TDEs -- and therefore the possibility of thermonuclear ignition -- has not reached a consensus \citep{Laguna1993,Brassart08,Gafton2019}. 
 
 Recently,  \citet{Norman21} and \citet[hereafter CN22]{Coughlin22a} analyzed the deep TDE regime using analytical and numerical methods, focusing mainly on a $\gamma = 5/3$ polytrope, and found that the $\beta^3$ scaling is generally not followed. However, they briefly considered a standard-model star \textit{analytically} and concluded that these stars also do not adhere to the above scaling. To further understand the compression experienced by a radiative star during a deep tidal encounter, here we numerically analyze the maximum central density and temperature achieved by a Sun-like star modeled with the Eddington standard model during a deep TDE. 
 
 In Section \ref{sec:analytic} we recapitulate the analytical analysis of CN22 adapted for standard-model stars. In Section \ref{sec:numerical} we present the results of numerical simulations, and we make comparisons to, and demonstrate excellent agreement with, the analytical model; we also analyze the convergence of the simulations with respect to particle number and briefly consider the effects of general relativity. We summarize and conclude in Section \ref{sec:summary}.

\section{Analytic Estimates}
\label{sec:analytic}
Here we summarize the analytical model first presented in CN22 and the predictions of that model for a standard-model star. We consider a star of mass $M_\star$ and radius $R_\star$, the pressure $p$ of which is related to the density $\rho$ according to $p \propto \rho^{4/3}$. We let the adiabatic index $\gamma$, defined such that $\gamma-1$ is the ratio of the pressure to the internal energy of the gas, be $\gamma = 5/3$, which is a very good approximation for low-mass stars. We assume that the star is in hydrostatic equilibrium far from the SMBH of mass $M_\bullet$, at which point the central density and pressure are $\rho_{\rm c}$ and $p_{\rm c}$, respectively. The corresponding scale length appropriate to the interior of the star is then 

\begin{equation}
    \alpha^2=\frac{6\gamma p_{\rm c}}{\rho_c} \frac{1}{4 \pi G \rho_{\rm c}},
\end{equation} 
where $G$ is the gravitational constant. We work in the tidal approximation, such that the center-of-mass motion is decoupled from the internal motions of the star, and we assume that the center-of-mass follows a parabolic orbit (i.e., the star is initially very far from the black hole where the velocity is $\sim 0$ compared to the velocity near pericenter). Then the distance of the center of mass $r_{\rm c}$ satisfies 

\begin{equation}
    \frac{1}{2} \left(\frac{\partial r_{\rm c}}{\partial t}\right)^2+\frac{G M_\bullet r_{\rm p}}{r^2_c}-\frac{GM_\bullet}{r_{\rm c}}=0,
    \label{eq:com}
\end{equation}
where $r_{\rm p}= r_{\rm t}/\beta$ is pericenter distance. The solution to Equation \eqref{eq:com} is (CN22)

\begin{equation}
    r_{\rm c}=r_{\rm p} \cosh^2(\tau),
    \label{eq:defrc}
\end{equation}
where $\tau$ is implicitly defined as

\begin{equation}
    \frac{\partial \tau}{\partial t}= \sqrt{\frac{G M_\bullet}{2 r^3_c}}.
    \label{eq:deftau}
\end{equation}
The pericenter is reached by definition at $\tau = t = 0$.
\begin{figure}
   \centering
    \includegraphics[width=0.47
    \textwidth]{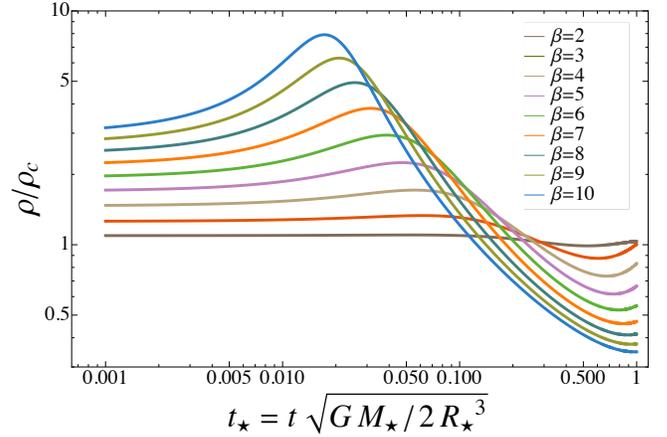}
     \caption{The ratio of the central density normalized to its original value as a function of time normalized by the dynamical time of the star. As $\beta$ increases the extent of compression increases significantly, and the time at which the maximum density is reached approaches $0$ (which is the time when the center of the star reaches the pericenter).}
    \label{fig:htau}
\end{figure}

\begin{figure*}
   \centering
    \includegraphics[width=0.48\textwidth]{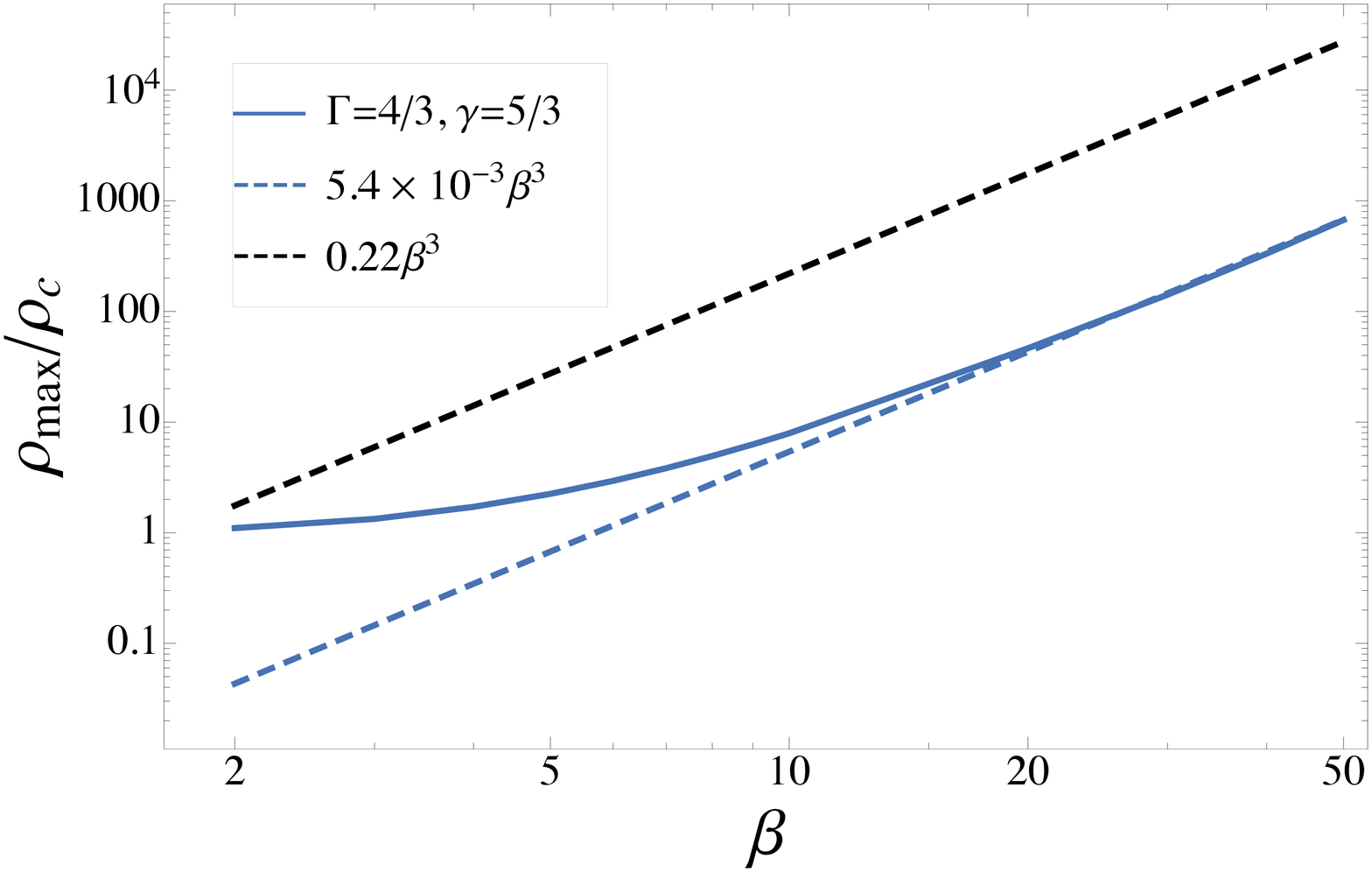}
    \includegraphics[width=0.48
    \textwidth]{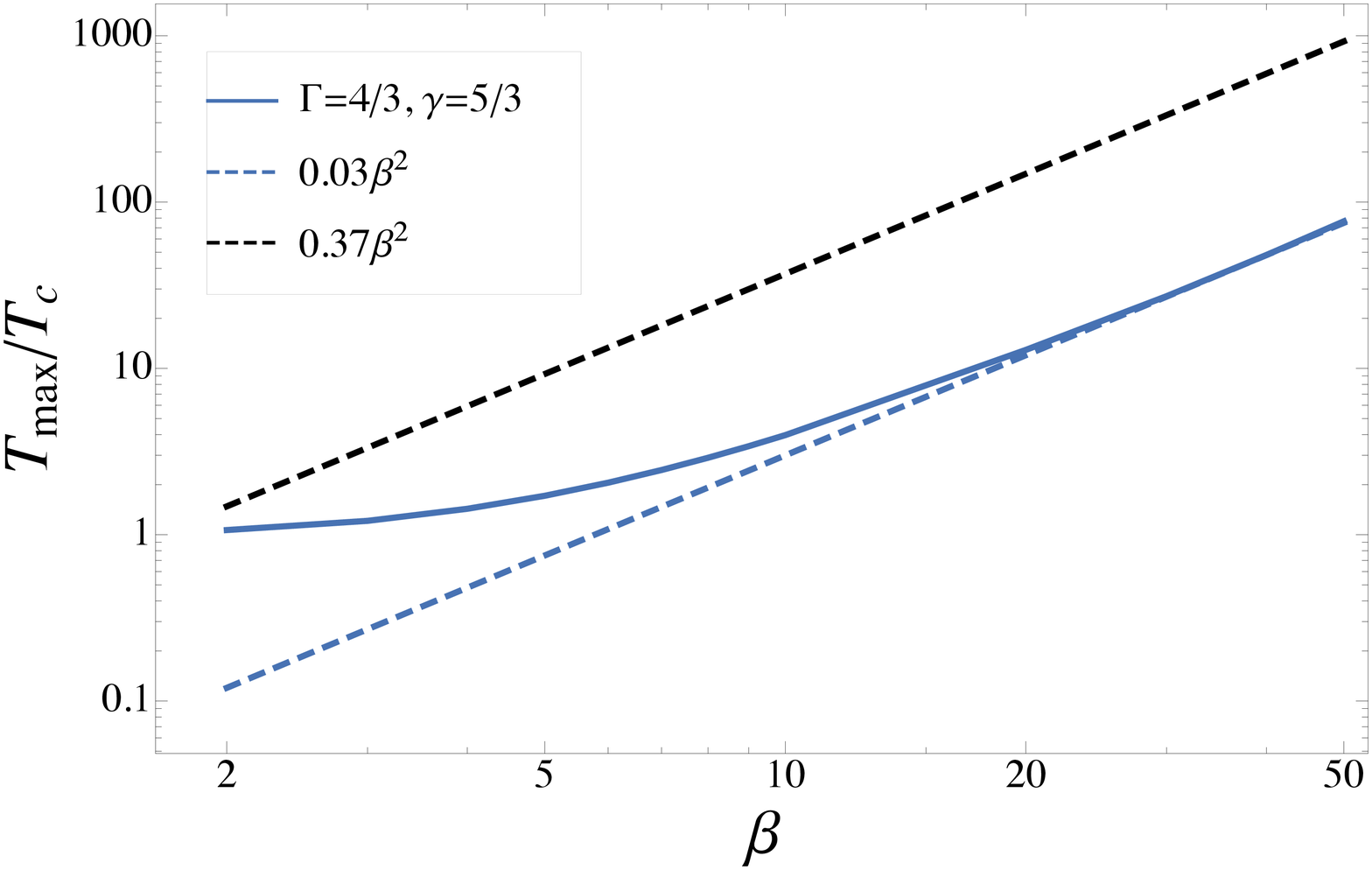}
     \caption{The maximum central density (left) and temperature (right) normalized to their original values as a function of $\beta$. The blue dashed lines indicate the large-$\beta$ behavior of our analytic solution, while the black dashed lines indicate the corresponding scaling predicted by \citet{Carter83, Luminet86}.}
    \label{fig:rhotvsbeta}
\end{figure*}
The fluid variables satisfy the momentum equations, the entropy equation, the continuity equation, and the Poisson equation. The model of CN22 proposes that, in the deep-TDE limit, the compression occurs predominantly in the $z$-direction, and hence we can approximately ignore the deviation in the in-plane motion (with respect to the center of mass). Adopting a homologous relationship between the height of a fluid element $z$ and its initial height $z_0$, 

\begin{equation}
    z= H(\tau) z_0,
    \label{eq:affine}
\end{equation}
we can show that the fluid equations can be combined into the following second-order differential equation for $H$:

\begin{equation}
    \mathcal{L}[H]-\frac{2}{\beta^3} \frac{\rho_{\rm c}}{\rho_{\star}} (H^{-\gamma}-1) \cosh^6(\tau)=0,
    \label{eq:operatoreom}
\end{equation}
where $\rho_{\star} = M_{\star}/(4\pi R_{\star}^3/3)$ is the average stellar density and the operator $\mathcal{L}$ is

\begin{equation}
    \mathcal{L}=\frac{\partial^2}{\partial \tau^2}-3 \tanh(\tau) \frac{\partial}{\partial \tau}+2
    \label{eq:dynop}.
\end{equation}
The terms after $\mathcal{L}[H]$ in Equation \eqref{eq:operatoreom} arise from pressure and self-gravity. Hydrostatic equilibrium at infinity implies $H(\tau \rightarrow - \infty)=1$ and $\dot{H}(\tau \rightarrow - \infty)=0$. The time-dependent central density is then 
\begin{equation}
    \rho(z,\tau)=\frac{\rho_{\rm c}}{H(\tau)}. 
    \label{eq:den}
\end{equation}

Figure \ref{fig:htau} shows the central density as a function of time normalized by the dynamical time of the star for the $\beta$ in the legend. We see that as $\beta$ increases, the maximum density obtained increases, and the time at which the maximum density is achieved occurs earlier (note that $t=0$ is when the center of mass reaches pericenter). The left (right) panel of Figure \ref{fig:rhotvsbeta} illustrates the maximum central density (temperature) as a function of $\beta$. Consistent with the assumption that the gas pressure dominates over radiation pressure, the temperature is calculated as $T \propto p/\rho$. The blue-dashed line in the left (right) panel shows the $\propto \beta^3$  $(\beta^2)$ fit to the large-$\beta$ behavior of our results. The maximum density (temperature) only begins to adhere to the $\beta^3 (\beta^2)$ scaling at $\beta \gtrsim 25$, much higher than the prediction of CL83, who reported that those scalings appear at $\beta \gtrsim 5$ (black dashed lines represent their prediction). 
Furthermore, the proportionality factors 
are more than an order of magnitude smaller than those given in \cite{Luminet86}.
As discussed in detail in CN22, the significantly reduced, maximum density (compared to the prediction of \citealt{Luminet86}) arises from the fact that the pressure gradient counteracts the tidal compression when the gas pressure is only a fraction of the free-falling ram pressure (see Figure 4 of CN22). 

\begin{figure*}
  \centering
    \includegraphics[width=0.99\textwidth]{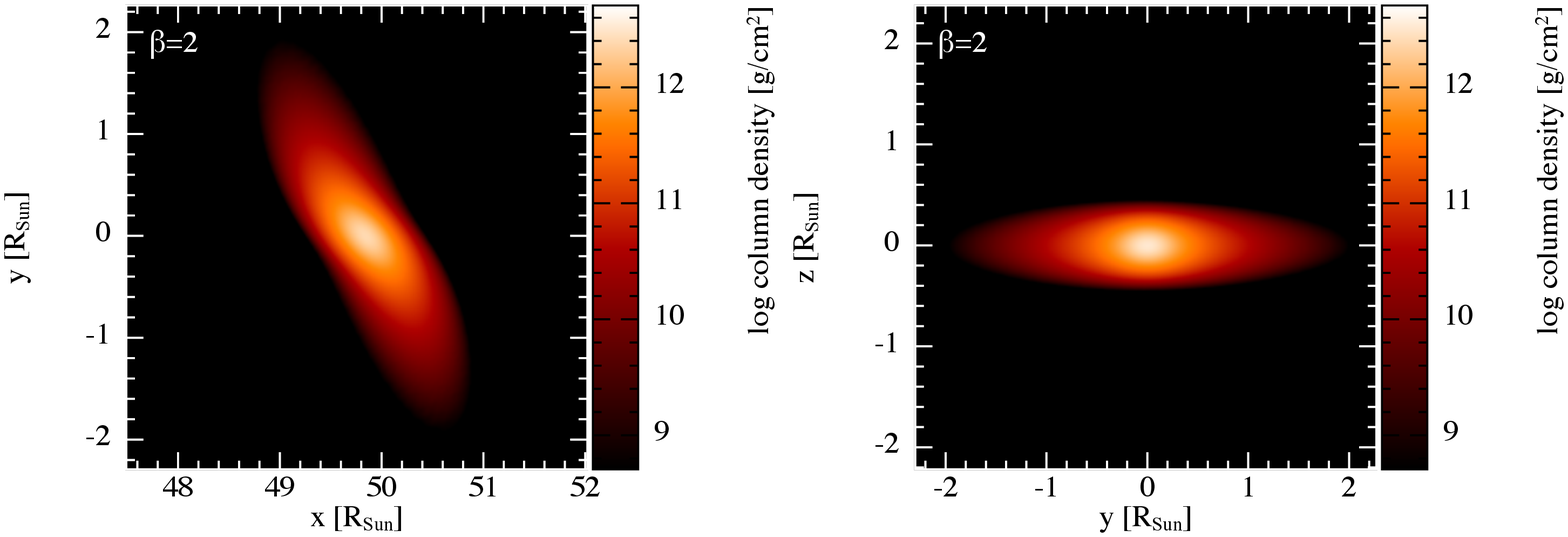}
    \includegraphics[width=0.99\textwidth]{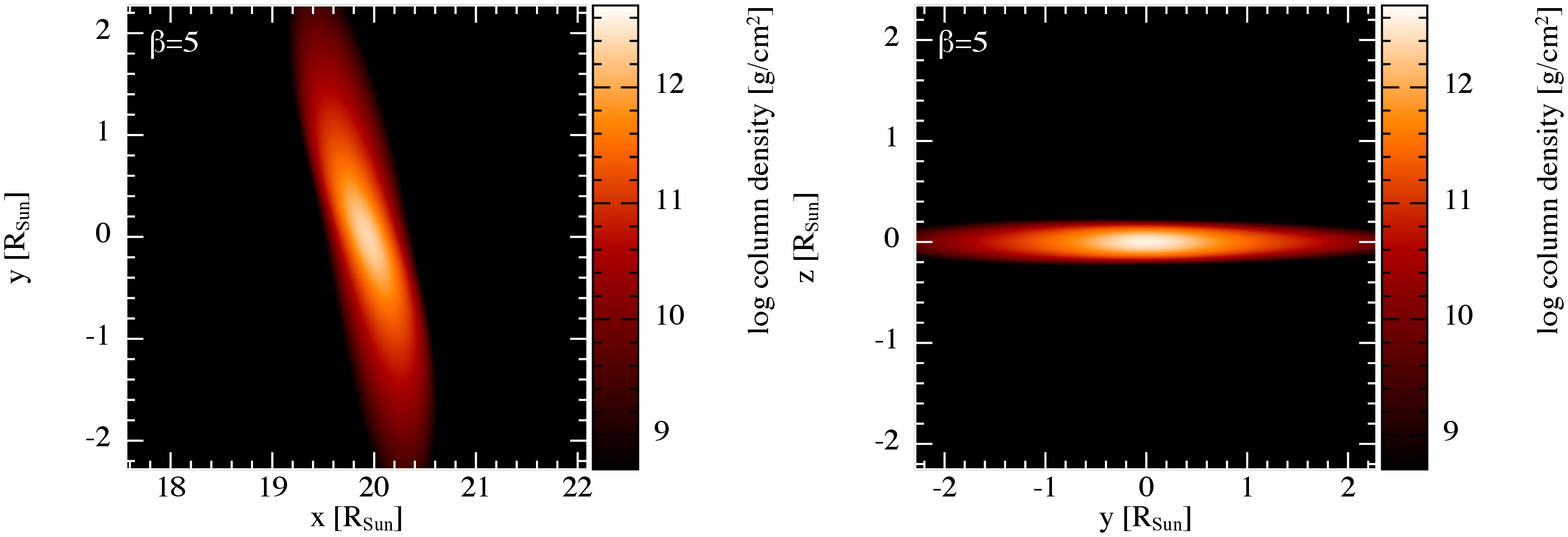}
    \includegraphics[width=0.99\textwidth]{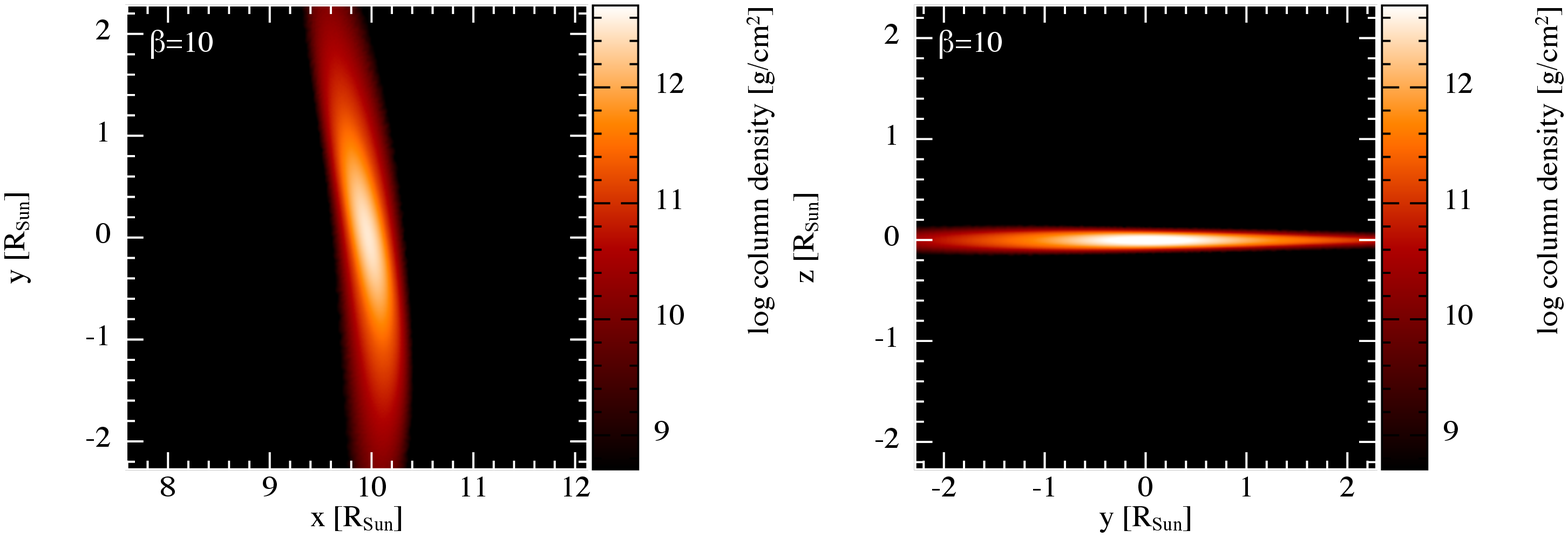}
     \caption{The integrated column density at the time when the center of mass of the star reaches pericenter. The value of $\beta$ is shown in the top-left of each panel. The left panels show the distribution of stellar material in the orbital plane, and the corresponding right panels show the view perpendicular to that plane. In the convention we adopt, both the initial $x,y$ coordinate of the center of mass of the star starts with negative values, and arrives the pericenter with a positive $x$-coordinate value and $y = 0$. Thus, any fluid element with a positive (negative) $y$-value has already (not yet) passed through its pericenter. Increasing $\beta$ clearly leads to an increase in the flattening, or ``crushing'' of the star, into the orbital plane near pericenter. }
\label{fig:columndensity}
\end{figure*}

The model presented here makes a number of approximations about the geometry of the system and the importance of vertical compression over in-plane motion. In the next section we analyze hydrodynamical simulations that relax these approximations to compare to these predictions. 

\section{Numerical simulations}
\label{sec:numerical}
\subsection{Simulation setup}
\label{sec:method} 
Here we present the results of numerical simulations of disruptions of Sun-like stars modeled with the Eddington standard model. We use the smoothed-particle hydrodynamics (SPH) code {\sc phantom} \citep{price18}, which has been widely used for studying TDEs \citep{Coughlin15,Coughlin16,Miles20,Norman21,Cufari21}.

The Eddington standard model is implemented in our code in the following way: a discrete radial grid is constructed with a large number of sufficiently close points extending from the center of the star to the surface. We then assign to these points the appropriate density and pressure that are obtained by numerically integrating the Lane-Emden equation. The configuration is then ``relaxed'' in isolation (i.e., without the gravitational field of the black hole) for ten sound crossing times to remove numerical perturbations. 
The center of mass of the relaxed star is then placed at a distance of $5r_{\rm t}$ from the black hole, so that all particles move with the center of mass, which is on a parabolic orbit with pericenter distance $r_{\rm p} = r_{\rm t}/\beta$. The self-gravity and viscosity switches are implemented through standard routines (see \citealt{Norman21}). We simulate encounters with $2\le \beta\le 10$ in integer steps.

\subsection{Simulation Results}
\label{sec:results}
In the left (right) panel of Figure \ref{fig:columndensity}, we present the integrated column density as seen in the orbital plane (out of the orbital plane), when the center of mass of the star reaches the pericenter. The $\beta$ of the encounter is shown in the top left of each panel. 
The pericenter is in the $x$-direction, the $x-y$ plane is the orbital plane, and the $y-z$ plane is orthogonal to the orbital plane. As seen in the figure, the star suffers a significant distortion in the process, and as $\beta$ increases, it is compressed vertically into a small fraction of its original volume. 
\begin{figure}
  \centering
    \includegraphics[width=0.47\textwidth]{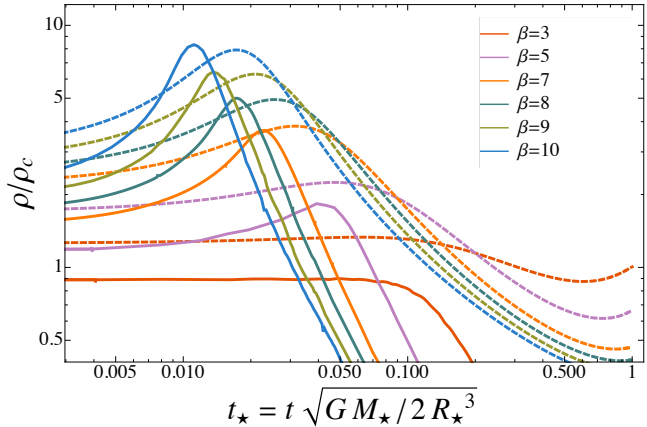}
     \caption{The central density normalized to its original value as a function of time normalized by the dynamical time of the star. The different curves are appropriate to the $\beta$ in the legend, with solid (dashed) curves resulting from the numerical simulations (analytic model). The magnitude of the compression and time of the maximum compression agrees reasonably well between the SPH and analytic results. }
\label{fig:rho_cmp_1M}
\end{figure}

\begin{figure*}
  \centering
    \includegraphics[width=0.495\textwidth]{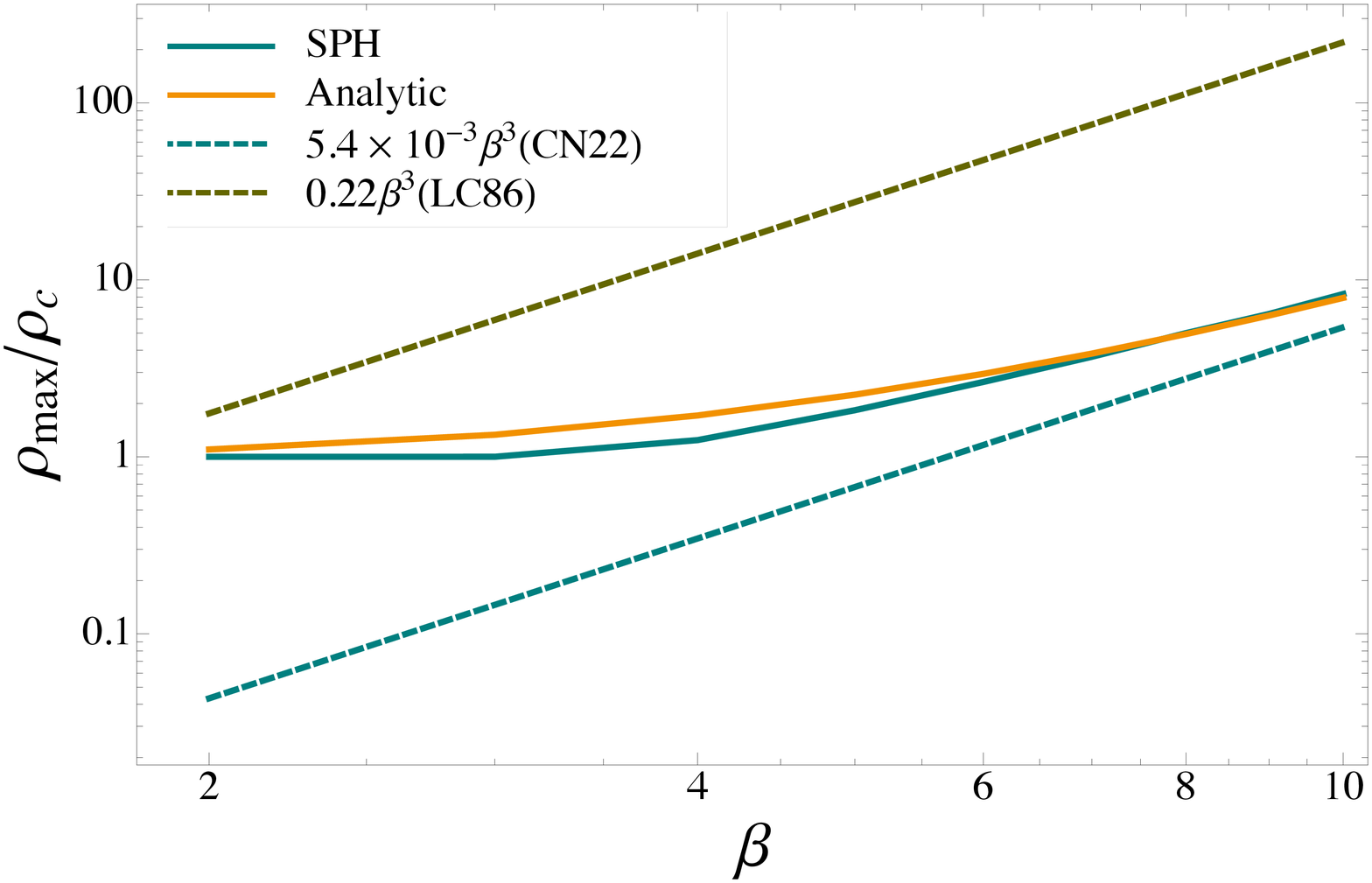}
     \includegraphics[width=0.495\textwidth]{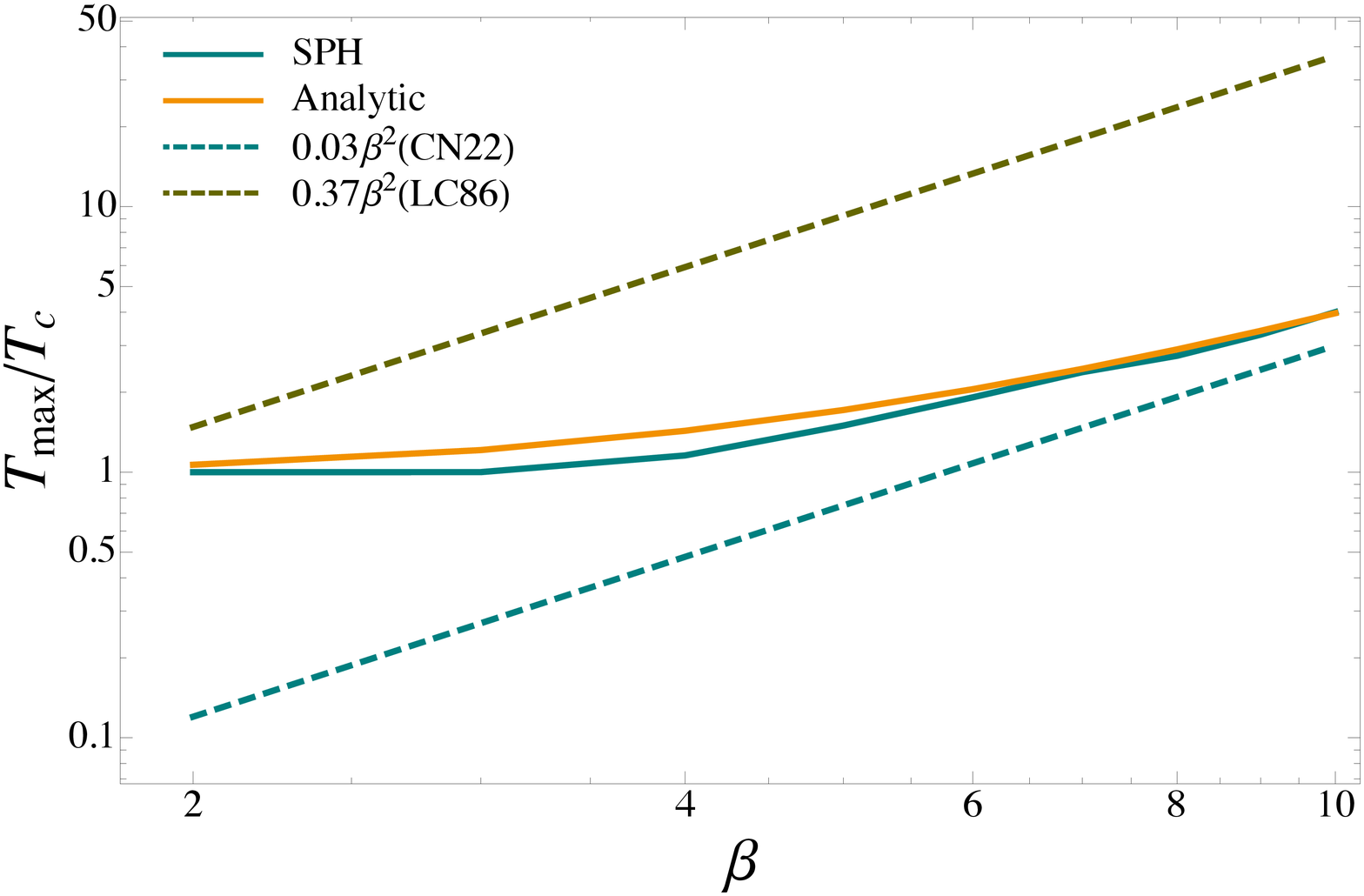}
    \caption{The maximum value of the central stellar density (left) and temperature (right) as a function of $\beta$ normalized by their initial values. The dashed curves represent the scalings derived by \citet{Luminet86} (olive) and \citet{Coughlin22a} (teal), the solid, teal curves are from the analytic model (also in \citealt{Coughlin22a}), and the solid, orange curves are from the numerical simulations.}
\label{fig:rho_cmp_ana}
\end{figure*}

\begin{figure}
  \centering
    \includegraphics[width=0.495\textwidth]{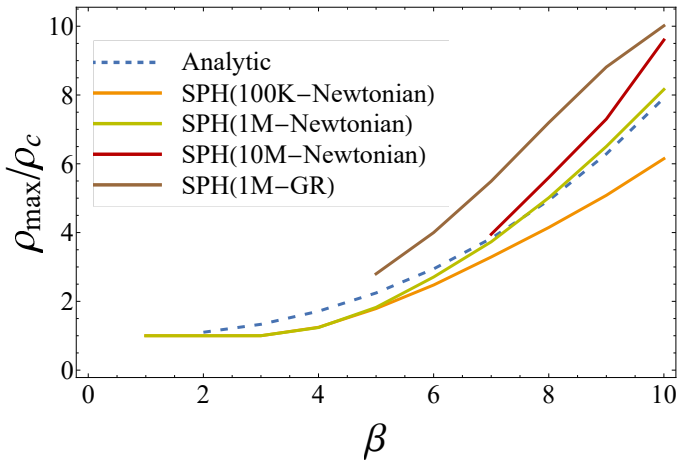}
     \caption{The ratio of the maximum to the original density at the geometric center of the star as a function of $\beta$. The different solid curves are obtained by varying the spatial resolution, with the corresponding number of particles shown in the legend. The analytic prediction is shown by the dashed curve.}
\label{fig:converg}
\end{figure}

Figure \ref{fig:rho_cmp_1M} shows the central density normalized to its original value as a function of time normalized by the dynamical time of the star. Solid curves result from simulations, whereas dashed curves are the corresponding analytic predictions. As $\beta$ increases, the magnitude of the maximum density achieved (that is, the height of the peak) during the encounter increases and the time at which the maximum density is achieved approaches zero. It is clear that the analytical and numerical results agree well in their prediction of the maximum density. The disagreement at other times is due to the fact that the homologous model presented here ignores both the in-plane stretching and nonlinear (i.e., non-homologous) effects, which approximately negate one another as concerns the maximum density. We also note that the star spends a very short fraction of its dynamical time near the maximum density (and correspondingly near the maximum temperature).

 In Figure \ref{fig:rho_cmp_ana} we show the maximum density (left) and temperature (right) against $\beta$ obtained from the simulations alongside the corresponding analytic prediction. For small values of $\beta$ the in-plane stretching is significant, and therefore the homologous model presented in Section \ref{sec:analytic}, which ignores any in-plane dynamics, does not account for the decrease in density and overestimates the numerical value; overall, however, the agreement between the analytic model and numerical solution is good. In each panel the $\beta^3$ fit to the large-$\beta$ behavior of our results is shown with a teal dashed curve and that of the LC86 with an olive dashed curve. In the $\beta$ range of our investigation, we find that neither the maximum central density nor the maximum temperature follow any power-law scaling, in contrast to the prediction of LC86 (who claimed the $\beta^3$ scaling appears for $\beta \gtrsim 5$). We tabulate the maximum density and temperature as obtained from the simulations, along with their predicted values from Section \ref{sec:analytic} (and the homologous model of CN22) and LC86, in Table \ref{table:comp} for the full $\beta$ range of our investigation. For $\beta = 1$ and 2, the in-plane stretching in the numerical simulation offsets the compression out of the plane, resulting in a monotonic decline in both the central density and temperature; hence these values are identically equal to 1.
 \begin{table}
 \centering
 \begin{tabular}{ccccccc}
  \hline
   & \multicolumn{3}{c}{$\rho_{\rm{max}}/\rho_c$}  &  \multicolumn{3}{c}{$T_{\rm {max}}/T_c$} \\
  \hline
   $\beta$ & SPH (1M) & LC86 & CN22 & SPH (1M) & LC86 & CN22 \\
  \hline
  2 & 1 & 1.76 & 1.10 & 1 & 1.48 & 1.07\\
  \hline
  3 & 1 & 5.94 & 1.33 & 1 & 3.33 & 1.2\\
  \hline
  4 & 1.24 & 14.08 & 1.71& 1.16 & 5.92 & 1.43\\
  \hline
  5 & 1.83 & 27.5 & 2.25 & 1.50 & 9.25 & 1.72\\
  \hline
  6 & 2.65 & 47.52 & 2.94 & 1.91 & 13.32 & 2.05\\
  \hline
  7 & 3.68 & 75.46 & 3.84 & 2.38 & 18.13 & 2.45\\
  \hline
  8 & 5.00 & 112.64 & 4.94 & 2.74 & 23.68 & 2.90\\
  \hline
  9 & 6.48 & 160.38 & 6.29 & 3.29 & 29.97 & 3.41 \\
  \hline
  10 & 8.18 & 220. & 7.91 & 4.01 & 37. & 3.97\\
    \hline
    \end{tabular}
    \caption{The maximum central density and temperature, normalized to their original values, obtained from the SPH simulations, predicted by LC86, and predicted by CN22 for the $\beta$ range analyzed here. For $\beta = 1$ and 2, the density and temperature at the center of the star monotonically decline with time in the numerical simulations, hence their values of identically 1 from the simulations.}
    \label{table:comp}
\end{table}

By extrapolating our simulation (or analytic) results, one could argue that the $\propto \beta^3$ scaling would hold at a much higher value of $\beta$, on the order of $\beta \simeq 20$, and with a much smaller proportionality factor than that of LC86. However, the analytical model here is at the homologous level and does not permit the formation of shocks, and while the agreement between the analytical model and the numerical simulations for $\beta \le 10$ suggests that shocks are not important over this $\beta$ range, they likely do become important for larger $\beta$. For example, CN22 demonstrated that, for a $\gamma = 5/3$ polytrope, the maximum density never actually conforms to the $\beta^3$ scaling (in any $\beta$ range) because a (weak) shock reaches the midplane prior to maximum adiabatic compression above $\beta \simeq 10$. A similar effect almost certainly occurs for this type of star as well, and hence it is likely inaccurate to extrapolate the homologous prediction and conclude that the $\beta^3$ scaling is eventually followed.
 
We tested the numerical accuracy of our results using three different resolutions, corresponding to $10^5$, $10^6$ and $10^7$ SPH particles, which are shown in Figure \ref{fig:converg} alongside the analytical results. It is clear that the simulations agree well with one another and the analytical results for small $\beta$, but disagree somewhat at large $\beta$ where the higher-resolution simulations predict a greater degree of compression. Nonetheless, it is apparent from this figure that the relative change in $\rho_{\rm max}/\rho_{\rm c}$ is a decreasing function of resolution, with the specific values given in Table \ref{table:err} (the \% change columns are calculated as the difference between the higher and lower-resolution values normalized to the high-resolution value). We therefore conclude that while the results have not definitively converged at $10^{7}$ particles for $\beta \gtrsim 8$, they are converging, and the amount of compression experienced by the star is an order of magnitude smaller than that predicted in previous works. We also note that a similar trend was found in \citet{Norman21}, where even at $10^{8}$ particles the results were not yet converged for $\beta \gtrsim 8$ (though they showed clear evidence that they were converging; see Figure 17 of \citealt{Coughlin22a}). Finally, while there is some disagreement between the $10^{7}$-particle runs and the homologous prediction for $\beta \gtrsim 8$, \citealt{Coughlin22a} have shown that incorporating non-homologous terms in the analytical solution can bring these two into better agreement (see Figure 17 of \citealt{Coughlin22a} for a demonstration of this in the case of a convective star).
\begin{table}
 \centering
 \begin{tabular}{@{}cccccc@{}}
  \hline
  \hline
  $\beta$  &  ${N}_{\rm p} = 10^{5}$ & ${N}_{\rm p} = 10^{6}$ & \% change & ${N}_{\rm p} = 10^{7}$ & \% change  \\
  \hline
  \hline
  ${7}$ & ${\rho_{\rm max}}/{\rho_{\rm c}} = 3.3$ & 3.7 & 12.1 & 3.8 & 2.7 \\
  \hline
  8 & 4.2 & 5.0 & 19.1 & 5.6 & 12\\
  \hline
  9 & 5.1 & 6.5 & 27.5 & 7.3 & 12.3 \\
  \hline
  10 & 6.2 & 8.2 & 32.3 & 9.6 & 17.1\\
  \hline
  
    \end{tabular}
     \caption{For the $\beta$ given in the first column, the maximum central density relative to its original value is given in columns 2, 3, and 5 for $10^5$, $10^6$ and $10^7$ SPH particles, respectively. The relative error between successive resolutions, calculated as the difference between the higher and lower-resolution values normalized by the higher-resolution result, is shown in the fourth and sixth column.}
    \label{table:err}
\end{table}

\subsection{Effects of General Relativity}
\label{sec:gr}
The analysis of Section \ref{sec:analytic} and the simulations presented so far have been performed in Newtonian gravity. This made the analysis simpler and the corresponding simulations computationally inexpensive. Furthermore, a Newtonian background has historically been preferred in almost all previous work investigating extreme tidal compression, specifically in 
CL82,83, LC86 and CN22 with which we compare our results. 

However, the pericenter distance of the star in units of gravitational radii for a $10^6M_{\odot}$ black hole is $r_{\rm p} \simeq 47/\beta$, and thus by $\beta = 10$ is very close to the direct capture radius (4 gravitational radii). Thus, general relativistic effects can modify the evolution of the compressing star nontrivially and, as argued in CN22, could increase the maximum-achieved density owing to the stronger tidal field of the black hole. To investigate this possibility, we performed general relativistic simulations in the (fixed-metric) Schwarzschild geometry using the SPH algorithm described in \cite{Liptai19}. The relativistic simulations were primarily performed using $10^6$ SPH particles for $\beta =5-10$ \footnote{Note that we are still defining $\beta$ by $\beta = r_{\rm t}/r_{\rm p}$, where $r_{\rm p}$ is the true pericenter distance the star would reach if it were a point particle in the relativistic gravitational field of the SMBH. We do not, in contrast, fix the angular momentum of the star to its Newtonian value and define $\ell^2 = 2GM_{\bullet}r_{\rm p}$, which would generally yield a smaller, true pericenter distance in the relativistic gravitational field of the SMBH; see \citep{Coughlin22b}.}, though an additional simulation with $10^{7}$ particles was performed for $\beta = 7$ to assess the convergence of the results.

The general relativistic results compared to the Newtonian values are shown by the brown curve in Figure \ref{fig:converg}. Compared to the green curve in this figure (which is at the same resolution), we see that general relativistic effects tend to increase the amount of compression by a factor of $\lesssim 1.5$. Interestingly, the relative change in the maximum-achieved density does not appear to be as pronounced for $\beta = 10$, which could be due to the fact that the direct capture radius for this configuration coincides with $\beta \simeq 11.8$. As the star nears the direct capture radius the tidal shear -- responsible for reducing the density of the material -- diverges, and one might therefore suspect that the overall degree of compression is reduced as the direct capture limit is reached. However, we make this interpretation with caution owing to the lack of complete convergence of the solutions for this value of $\beta$.

\begin{figure}
  \centering
    \includegraphics[width=0.45\textwidth]{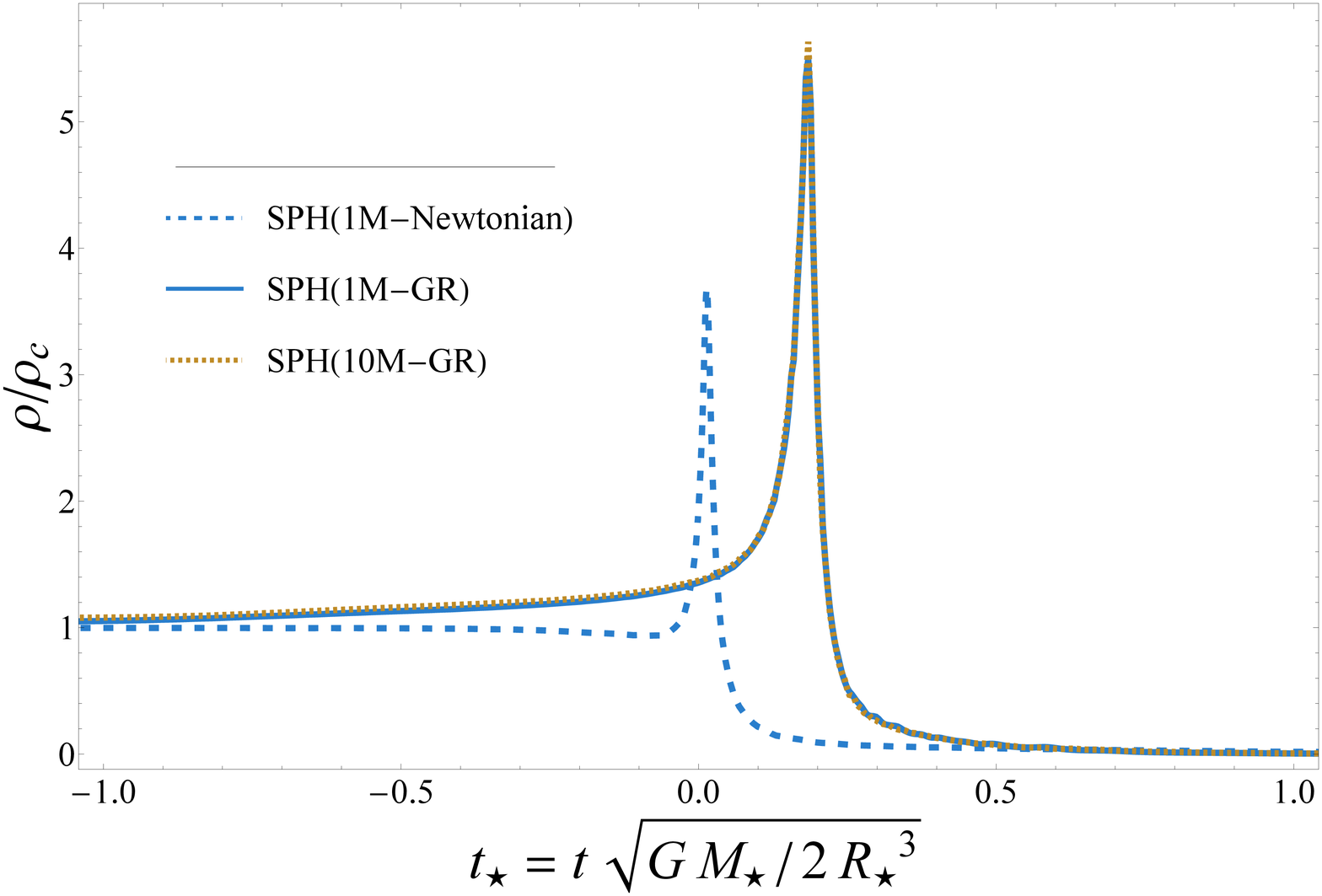}
     \caption{The ratio of maximum central density to the original central density against the dynamic time of the star with  $\beta=7$. The resolution and gravity are specified by the legend (solid-blue and dotted-orange use the Schwarzschild metric for the gravitational field of the SMBH). The compression is stronger (by a factor of $\lesssim 1.5$) with relativistic effects included, and the time of maximum compression occurs later in comparison to the Newtonian case; the latter effect arises from time dilation (note that the horizontal axis is coordinate time, i.e., time as measured by an observer at infinity) normalized by the dynamic time of the star.}
\label{fig:gr}
\end{figure}

Figure \ref{fig:gr} shows the central density as a function of coordinate time for the $N_{\rm p} = 10^{6}$ Newtonian, $N_{\rm p} = 10^{6}$ relativistic, and $N_{\rm p} = 10^{7}$ relativistic simulations for $\beta = 7$. Consistent with Figure \ref{fig:converg}, the maximum density attained in the relativistic simulations is $\lesssim 1.5$ times the Newtonian value. Additionally, the time at which the star is maximally compressed is slightly delayed (note that the horizontal axis is coordinate time relative to when the Newtonian, point-particle orbit would reach pericenter), and the overall duration of the compression (i.e., the amount of time that the star spends near its maximum-achieved density) is prolonged \footnote{We thank Emilio Tejeda for pointing out this latter feature of the relativistic solutions.} in the general relativistic solutions compared to the Newtonian one. Both of these effects arise from relativistic time dilation. It is also evident that the results of the relativistic simulations with $10^6$ and $10^7$ SPH particles agree extremely well with one another.

\section{Summary and Conclusions}
\label{sec:summary}
To analyze the amount of tidal compression of a radiative, solar-like star (modeled with the Eddington standard model) during a tidal disruption event, we used an analytic model originally proposed by \citet{Coughlin22a} that accounts for both the self-gravity and the pressure of the star during its tidal encounter with the black hole (Section \ref{sec:analytic}). We then relaxed the assumptions made within that model by performing three-dimensional simulations of deep TDEs that satisfied $2 \le \beta \le 10$, where $\beta = r_{\rm t}/r_{\rm p}$ with $r_{\rm p}$ the stellar pericenter distance and $r_{\rm t}$ the canonical tidal radius, verified the numerical accuracy of our results by varying the spatial resolution, and performed additional simulations in the Schwarzschild metric to assess the importance of general relativity (Section \ref{sec:numerical}). 
We showed that the two methods -- analytical and numerical -- agree very well in their predictions for the maximum density and temperature reached during the disruption, and we therefore conclude that

\begin{enumerate}
    \item The maximum density and temperature achieved by the star during its compression are significantly reduced (by over an order of magnitude for the density and nearly an order of magnitude for the temperature by $\beta = 10$) compared to the predictions of \citet{Carter1982}, \citet{Carter83}, \citet{Carter85}, \citet{Luminet86} (see Figure \ref{fig:rho_cmp_ana}).
    \item Shocks are not important in this range of $\beta$, and therefore are not responsible for the lower degree of compression, although they are likely important for sufficiently large $\beta$ ($\beta \gtrsim 10$; cf.~\citealt{Coughlin22a}).
    \item The predicted scaling $\rho_{\rm max} \propto \beta^3$ \citep{Carter1982, Brassart08, Stone2013} is not realized over this range of $\beta$, and is lik
    ely not ever followed because of the eventual importance of shocks in reversing the compression of the star prior to reaching its maximum adiabatic value (see \citealt{Coughlin22a}).
    \item General relativity modestly increases the maximum degree of compression of the star (by a factor of $\lesssim 1.5$; see Figures \ref{fig:converg} and \ref{fig:gr} and note that the general relativistic solutions are at $10^6$ particles), and also induces a lag in the time at which the maximum compression occurs and the amount of time the star spends at increased density and temperature increases (according to an observer at infinity) as a result of time dilation.
    \item The high temperatures and densities needed to ignite the triple-$\alpha$ process in the core of the star are not reached by $\beta = 10$, as the maximum temperature attained at this $\beta$ is $T_{\rm max} \simeq 4 \times 10^7$ for an initial central temperature of $10^{7}$ K (see right panel of Figure \ref{fig:rho_cmp_ana}), and in general we expect the nuclear energy released to be minimal because of the small amount of time spent near maximum compression (see Figure \ref{fig:rho_cmp_1M}). Nevertheless, this modest degree of compression could still be important for augmenting the importance of self-gravity in the compressing star and thus determining the critical $\beta$ at which the star is completely destroyed, particularly for more massive stars where the critical $\beta$ is $\gtrsim 3$ \citep{law-smith20, Coughlin22c}.
\end{enumerate}.

\section*{Acknowledgements}
We thank the referee, Emilio Tejeda, for useful and constructive comments. S.K.K.~and E.R.C.~acknowledge support from the National Science Foundation through grant AST-2006684, and E.R.C.~acknowledges additional support from the Oakridge Associated Universities through a Ralph E.~Powe Junior Faculty Enhancement Award. C.J.N.~acknowledges support from the Science and Technology Facilities Council [grant number ST/W000857/1]. Some of this work was carried out using the Syracuse University HTC Campus Grid and the NSF award ACI-1341006. We used SPLASH \citep{Price07} for Figure \ref{fig:columndensity}.
\bibliography{highbeta.bib}{}
\bibliographystyle{aasjournal}

\end{document}